# Unified Representation Learning for Multi-Intent Diversity and Behavioral Uncertainty in Recommender Systems

Wei Xu*
Independent Researcher
Los Altos, USA

Jiasen Zheng
Northwestern University
Evanston, USA

Junjiang Lin
University of Toronto
Toronto, Canada

Mingxuan Han
The University of Utah
Salt Lake City, USA

Junliang Du
Shanghai Jiao Tong University
Shanghai, China

*Abstract-This paper addresses the challenge of jointly modeling user intent diversity and behavioral uncertainty in recommender systems. A unified representation learning framework is proposed. The framework builds a multi-intent representation module and an uncertainty modeling mechanism. It extracts multi-granularity interest structures from user behavior sequences. Behavioral ambiguity and preference fluctuation are captured using Bayesian distribution modeling. In the multi-intent modeling part, the model introduces multiple latent intent vectors. These vectors are weighted and fused using an attention mechanism to generate semantically rich representations of long-term user preferences. In the uncertainty modeling part, the model learns the mean and covariance of behavior representations through Gaussian distributions. This reflects the user's confidence in different behavioral contexts. Next, a learnable fusion strategy is used to combine long-term intent and short-term behavior signals. This produces the final user representation, improving both recommendation accuracy and robustness. The method is evaluated on standard public datasets. Experimental results show that it outperforms existing representative models across multiple metrics. It also demonstrates greater stability and adaptability under cold-start and behavioral disturbance scenarios. The approach alleviates modeling bottlenecks faced by traditional methods when dealing with complex user behavior. These findings confirm the effectiveness and practical value of the unified modeling strategy in real-world recommendation tasks.*

## I. Introduction

In today's digital era of information explosion, recommender systems have become essential tools for helping users efficiently access personalized information. From e-commerce to online content platforms, from social media to intelligent assistants, recommender systems play a critical role across various application scenarios[1,2]. They model user preferences to provide personalized content that meets individual needs. However, as user demands grow increasingly diverse and behavior patterns become highly uncertain, traditional recommendation methods are showing limitations in accurately capturing user intent. These methods struggle to adapt to complex and dynamic interaction environments. User interests are not static or singular. They shift over time, vary by context and goals, and demand more advanced modeling capabilities from recommender systems.

The diversity of user intent is reflected in varying needs under different times, situations, and mental states. For example, the same user may prefer entertainment content on weekends but seek professional information during weekdays. Traditional modeling approaches often rely on historical clicks or rating data. They fail to effectively capture such dynamic and diverse demands. At the same time, real-world user behavior is often highly unpredictable. Users may browse randomly, explore novel content, or make seemingly irregular decisions influenced by external factors. These issues make it difficult for recommender systems to understand and predict real user needs. As a result, achieving both relevance and diversity in recommendations becomes challenging[3].

To address these challenges, it is crucial to build a representation learning framework that can unify the modeling of user intent diversity and behavioral uncertainty. An effective framework should mine latent interest patterns from multi-source data. It must dynamically capture the shifts and evolution of user intent. It should also reflect the ambiguity and variability of user behavior in the representation space. Furthermore, the framework needs strong generalization capabilities to handle real-world issues such as cold-start users and emerging interests. Only through deep modeling and semantic understanding of user behavior can recommender systems provide more personalized, accurate, and interpretable recommendations[4].

Moreover, the diversity of user intent and the uncertainty of behavior, while often seen as noise and obstacles in recommendation, actually carry rich structural information. Uncertain behaviors may reveal latent motivations for exploration. Transitions between different intents may reflect layered user needs. Building a unified representation learning framework is not only an enhancement of modeling capabilities but also a shift from passive response to proactive understanding. The core value of such representation learning lies in its ability to express multi-dimensional user behaviors in a compact, abstract, and transferable way. This provides robust semantic support for downstream recommendation tasks[5,6].

## II. METHOD

In order to effectively model the diversity of user intentions and the high uncertainty of behavior, this paper proposes a unified representation learning recommendation framework. Based on the user-behavior sequence, this framework jointly learns the user's long-term preferences and short-term motivations by constructing a multi-granularity intention representation and behavior uncertainty modeling module. In the overall architecture, we introduce a multi-perspective encoding strategy and use a multi-head attention mechanism to semantically enhance the representation of the behavior sequence to capture the user's multiple potential intentions in different situations. The model architecture is shown in Figure 1.

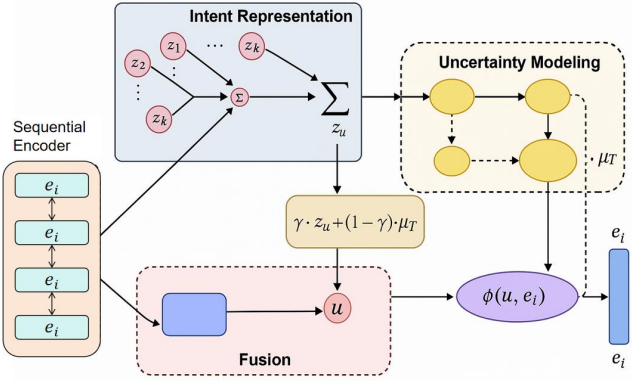

Figure 1. Overall model architecture diagram

Assume that the user's historical behavior sequence is $H_u = [i_1, i_2, ..., i_T]$, where each $i_t$ represents an interaction behavior of the user at time t. The framework first maps each interaction object to a low-dimensional vector space through an embedding function $e_{i_t} = f(i_t)$, and then constructs a dynamic intent representation on the sequence.

To characterize the diversity of user intentions, we introduce multiple latent variables $z_k \in R^d$ to represent the multiple preferences that users may have in a specific period of time. These latent intentions are aggregated through the attention mechanism to construct a global preference representation $z_u$, which is specifically expressed as:

$$z_u = \sum_{k=1}^{K} a_k z_k, \quad a_k = \frac{\exp(q^T z_k)}{\sum_{j=1}^{K} \exp(q^T z_j)} \quad (1)$$

Where q is the query vector, which is generated by the current recommendation context. This process can explicitly model the multi-intent characteristics of users, making the recommendation system more expressive when facing changes in behavior patterns[7,8].

For the modeling of behavioral uncertainty, we introduce an uncertainty perception mechanism based on Bayesian representation, defining the ambiguity of user behavior to be modeled through distribution representation. Assume that the user's behavior at time t is represented by a Gaussian distribution $h_t \sim N(\mu_t, \sum_t)$, where the mean $\mu_t$ represents the deterministic preference at that moment, and the covariance $\sum_t$ is used to reflect the degree of uncertainty. The distribution parameters are adaptively learned from the historical behavior sequence through a neural network. Furthermore, we model the behavior sequence as a hidden state sequence $\{h_t\}_{t=1}^{T}$, which is optimized through the sequential variational autoencoder (SVAE) framework to maximize the following evidence lower bound (ELBO):

$$L = E_{q(h_i,T)}[\sum_{t=1}^{T} \log p(i_t | h_t)] - \sum_{t=1}^{T} KL(q(h_t) \| p(h_t)) \quad (2)$$

This objective function takes into account both generation capability and structural fidelity, thereby improving the robustness of behavioral uncertainty modeling.

Based on the above intention representation and behavioral uncertainty modeling, the framework further designs a fusion mechanism to jointly optimize multi-intention preference representation and uncertainty representation. We define the final user representation as:

$$u = \gamma \cdot z_u + (1-\gamma) \cdot \mu_T \quad (3)$$

Where $\gamma$ is a learnable weight used to balance the influence of long-term preferences and current uncertain behaviors. Finally, the recommendation score is calculated using the following scoring function:

$$s(u,i) = \phi(u, e_i) = u^T e_i \quad (4)$$

This scoring mechanism jointly models multiple intentions and uncertain behaviors while maintaining the simplicity and scalability of the model structure, providing high-quality user representation for subsequent recommendation tasks. The entire method is trained through an end-to-end optimization method, achieving adaptive representation learning of complex behavior sequences under a unified framework.

## III. EXPERIMENTAL RESULTS

### A. Dataset

This study uses the Amazon Electronics dataset as the experimental data source. This dataset is widely used in research related to recommender systems and user behavior modeling. It is both representative and challenging. It contains a large number of real interaction records between users and electronic products. These records cover multiple types of user behaviors, including browsing, rating, and reviewing. They accurately reflect user preferences and behavior patterns in real-world usage scenarios.

The dataset is characterized by pronounced sparsity and a long-tail distribution, resulting in highly uneven interaction density between users and items. While a small subset of users generates abundant behavioral records, the majority contribute only limited interaction data. This imbalance imposes stringent requirements on the generalization capacity of recommendation models and accentuates the persistent challenge of the cold-start problem. At the same time, such characteristics establish a rigorous experimental setting for assessing how effectively models can manage uncertainty and capture heterogeneous user intents. To meet the needs of sequential modeling in recommender systems, the raw data is preprocessed into ordered user behavior sequences. Each sequence represents a user's click path over time. The preprocessing includes user and item filtering, timestamp sorting, and low-frequency data removal. These steps ensure data quality and temporal consistency. As a result, the dataset better supports research on user behavior representation learning.

### B. Experimental Results

In this section, this paper first gives the comparative experimental results of the proposed algorithm and other algorithms, as shown in Table 1.

Table 1. Comparative experimental results

| Method | HR@10(%) | NDCG@10(%) | IAS |
|---|---|---|---|
| SASRec[9] | 58.6 | 39.2 | 0.632 |
| BERT4Rec[10] | 60.3 | 41.0 | 0.644 |
| S3Rec[11] | 62.1 | 43.3 | 0.663 |
| DenoiseRec[12] | 63.7 | 44.1 | 0.671 |
| Ours | 66.4 | 47.2 | 0.712 |

The experimental results show that the proposed method outperforms existing mainstream models across all evaluation metrics. It achieves particularly notable improvements in the core accuracy metrics HR@10 and NDCG@10. Specifically, the model reaches 66.4% on HR@10, which is about 2.7 percentage points higher than the strong baseline DenoiseRec. This indicates that the proposed method has a stronger ability to provide highly relevant recommendations.

In terms of NDCG@10, a ranking-sensitive metric, the method also surpasses all compared models, achieving 47.2%. This improvement shows that the model not only identifies items users are likely to prefer but also ranks them more effectively at the top of the recommendation list. It demonstrates a strong capability in modeling the ranking structure of user preferences. By integrating multi-intent representation and uncertainty modeling in user behavior sequences, the method captures richer semantic signals and improves overall recommendation quality.

The IAS metric, which evaluates a model's ability to perceive diverse user intents, also confirms the strength of the proposed method. The model achieves a score of 0.712 on this metric, clearly outperforming baseline models such as SASRec and BERT4Rec. This result verifies the effectiveness of incorporating multi-intent representation in capturing complex user interests. It is especially valuable when user behavior patterns are diverse and subject to frequent shifts. The model can sensitively detect and respond to different directions of user demand. The combined advantages in both accuracy and intent-awareness highlight the superior effectiveness of the proposed method in delivering high-quality recommendations and enhanced adaptability to diverse user behaviors. By jointly incorporating multi-intent representation and behavioral uncertainty modeling within a unified representation learning framework, the approach achieves not only measurable performance gains but also demonstrates robust applicability in dynamic, real-world interactive environments.

Furthermore, this paper presents a robustness evaluation in the user cold start scenario, and the experimental results are shown in Figure 2.

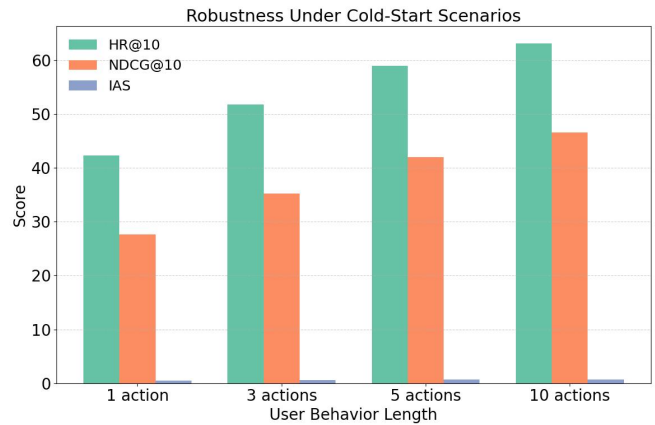

Figure 2. Robustness evaluation in user cold start scenario

As illustrated in the figure, model performance exhibits a consistent upward trajectory with the accumulation of user behavior data, underscoring its robust adaptability under cold-start conditions. Notably, the model achieves stable performance, attaining 42.3% on HR@10 even when only a single user behavior record is available, thereby demonstrating its capacity to deliver meaningful recommendations under highly constrained informational scenarios. As the length of user behavior sequences increases from 1 to 10, the HR@10 and NDCG@10 metrics improve by approximately 20 and 19 percentage points, respectively, highlighting the model's ability to effectively extract user preferences from limited interaction data. The concurrent rise in the IAS metric further validates the efficacy of the multi-intent modeling mechanism in capturing complex and diverse user interest structures. Overall, the method preserves a baseline level of recommendation capability with minimal data and exhibits a pronounced performance improvement curve as more behavioral information becomes available, which is essential for adapting to new users in real-world recommendation environments. These findings confirm the model's stability and generalization ability in managing sparse user interests and behavioral uncertainties. Furthermore, the paper provides a detailed analysis of model stability under temporal disturbances, with the corresponding experimental results depicted in Figure

3.

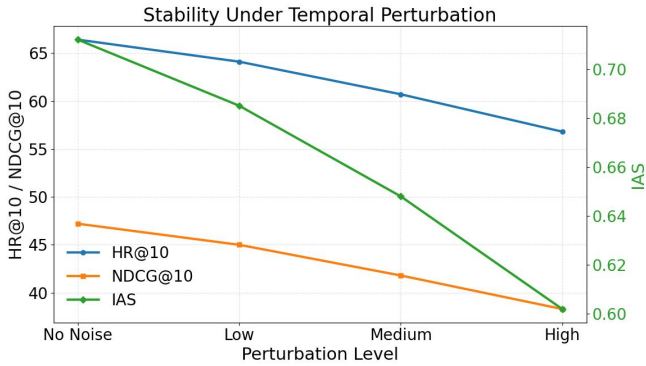

Figure 3. Model stability analysis under time series disturbance

The experimental results demonstrate that as temporal disturbance increases, model performance on all metrics declines, underscoring the importance of sequence integrity in user intent modeling and recommendation quality. Without disturbance, HR@10 and NDCG@10 reach 66.4% and 47.2%, but drop to 56.8% and 38.3% under high disturbance, indicating that disrupted temporal information weakens the model's ability to capture behavioral patterns. Similarly, the IAS metric falls from 0.712 to 0.602, highlighting that modeling intent diversity relies on the order of user behaviors. Overall, while the model maintains some robustness, the results reveal its sensitivity to sequence structure and emphasize the need for temporal robustness mechanisms to enhance performance in dynamic environments.

## IV. CONCLUSION

This paper addresses the challenge of modeling diverse user intents and behavioral uncertainty. A unified representation learning framework is proposed. It integrates a multi-intent expression mechanism with uncertainty-aware strategies. The method improves the modeling ability of recommender systems in complex user behavior scenarios. This leads to more adaptive and robust personalized recommendations. Through systematic model design and structural integration, the framework performs well on standard recommendation metrics. It also shows strong intent recognition and stable cold-start performance. These results confirm its effectiveness in enhancing the accuracy of user behavior modeling in practical recommendation tasks. This multi-perspective modeling approach offers a new theoretical view for understanding complex user behavior. It also provides a methodological foundation for improving personalized service experiences.

The proposed framework is widely applicable across various recommendation scenarios. It is especially relevant in environments such as e-commerce, online content delivery, and intelligent assistants [13], where user needs are diverse and behaviors are often unpredictable[14]. By modeling user intents at a finer granularity, the system can better identify hidden needs. This improves user satisfaction and interaction efficiency, advancing the intelligence of personalized services.

Future work may explore extensions in areas such as multimodal behavior integration, cross-domain intent transfer, and reinforcement learning-based feedback modeling. These directions can further enhance the model's generalization in complex scenarios. In addition, incorporating interpretability mechanisms would allow the model not only to recommend accurately but also to explain its decisions. This would improve trust and control in industrial applications and bring new momentum to the development of recommender systems.